# Layer Dependent Thermal Transport Properties of One- to Three-Layer Magnetic Fe:MoS$_2$

**Elham Easy**[1], **Mengqi Fang**[1], **Mingxing Li**[2], **Eui-Hyeok Yang**[1], **and Xian Zhang**[1,*]

[1]Stevens Institute of Technology, Department of Mechanical Engineering, Hoboken, New Jersey 07030, USA

[2]Brookhaven National Laboratory, Center for Functional Nanomaterials, Upton, New York 11973, USA

[*]xzhang4@stevens.edu

## ABSTRACT

Two-Dimensional (2D) transition metal dichalcogenides (TMDs) have been the subject of extensive attention thanks to their unique properties and atomically thin structure. Because of its unprecedented room-temperature magnetic properties, iron-doped MoS$_2$ (Fe:MoS$_2$) is considered the next-generation quantum and magnetic material. It is essential to understand Fe:MoS$_2$'s thermal behavior since temperature and thermal load/activation are crucial for their magnetic properties and the current nano and quantum devices have been severely limited by thermal management. In this work, Fe:MoS$_2$ is synthesized by doping Fe atoms into MoS$_2$ using the chemical vapor deposition (CVD) synthesis and a refined version of opto-thermal Raman technique is used to study the thermal transport properties of Fe:MoS$_2$ in the forms of single (1L), bilayer (2L), and tri-layer (3L). In the Opto-thermal Raman technique, a laser is focused on the center of a thin film and used to measure the peak position of a Raman-active mode. The lateral thermal conductivity of 1-3L of Fe:MoS$_2$ and the interfacial thermal conductance between Fe:MoS$_2$ and the substrate were obtained by analyzing the temperature-dependent and power-dependent Raman measurement, laser power absorption coefficient, and laser spot sizes. At the room temperature, the lateral thermal conductivity of 1-3L Fe:MoS$_2$ were discovered as 24 ± 11, 18 ± 9, and 16 ± 8 W/m·K, respectively which presents a decreasing trend from 1L to 3L and is about 40% lower than that of MoS$_2$. The interfacial thermal conductance of between Fe:MoS$_2$ and the substrate were discovered to be 0.3 ± 0.2, 1.1 ± 0.7, and 3.0 ± 2.3 MW/m$^2$·K for 1L to 3L respectively. We also characterized Fe:MoS$_2$'s thermal transport at high temperature, and calculated Fe:MoS$_2$'s thermal transport by density theory function. These findings will shed light on the thermal management and thermoelectric designs for Fe:MoS$_2$ based nano and quantum electronic devices.

## Introduction

Two-dimensional (2D) materials and their unique properties have attracted significant attention in recent years. With the discovery of graphene[1], transition metal dichalcogenides (TMDCs)[2-3] have been researched extensively due to their interesting thermal[4-8], electrical[9], thermoelectric[10], and optical[11-12] properties. Among them, $MoS_2$ is a layered semiconductor composed of covalently bonded transition-metal atoms (Mo) and chalcogen atoms (S)[13] with important applications in optoelectronic[14] and nano-electronic devices[15]. The thermal transport properties of 2D $MoS_2$ are very different from the bulk. Due to higher efficiency, thin layer $MoS_2$ are promising electronic devices[16]. Impurities, such as vacancies, chemical doping, and grain boundaries, can lead to phonon scattering[17-20] and, as a result, thermal conductivity reduction. Green's function suggests the higher likeliness of short wavelength or high-frequency phonon modes scattering by defects. Jenisha et al.[14] revealed the reduction of thermal conductivity for Ni-doped $MoS_2$. They observed two times smaller thermal conductivity of about 0.48 W/m·K for 7.5% of Ni-doped $MoS_2$ than undoped $MoS_2$. Mao et al.[15] reported a significant increase in electric conductivity and reduction in thermal conductivity for 1-10% vanadium atoms doped in $MoS_2$ lattice structure. Also, the thermal conductivity was significantly reduced due to point defects and interface scattering. Structural, optical, electrical, and phonon properties of V-doped and Ti-doped 2D $MoS_2$ were studied by Williamson et al.[21] They revealed the concentration of 2.083% Ti is the most effective due to effective dopant isolation. The research indicated that Ti dopant impacts the phonon properties greater than V-doped; however, V-dopant has a great phonon anharmonicity due to shorter bond length and weaker covalent bonds.

Modifying the magnetic properties of $MoS_2$ nanosheets is crucial for expanding their applications in nanoelectronics[22] and spintronics[23] using the diluted magnetic semiconductor (DMS) concept. $MoS_2$ nanosheets, however, are intrinsically nonmagnetic. Therefore, it is highly desirable to realize stable magnetism in $MoS_2$. Iron-doped Molybdenum disulfide (Fe:$MoS_2$) is considered a quantum material thanks to its intriguing properties, such as magnetic properties, lower mobility, more lattice impurities, and imperfections. Iron (Fe) doping into $MoS_2$ lattice structure has been reported to facilitate DMS formation[24], with Scanning Transmission Electron Microscopy (STEM), X-ray Photoelectron Spectroscopy (XPS), and Magnetic circular Dchroism (MCD) Spectroscopy characterizations of the Fe doping. The magnetic and electronic properties of Fe-doping monolayer $MoS_2$ have been studied using a combination of density functional theory by Shu et al.[25] They reported that the stability of Fe dopant in $MoS_2$ lattice structure depends on the layer number and chemical potential. In another work, Mishra et al.[26] investigated the electronic and magnetic properties of the Fe atom doped on the $MoS_2$ and $MoS_2$ using first-principal function theory. They reported that the dopant significantly changes magnetic moments, magnetic anisotropic energy, and charge transfer which counts as electronic properties. Wang et al.[11] compared undoped and Fe-doped $MoS_2$ chemical vapor deposition (CVD) grown optical and electrical properties. They observed longer carrier lifetime and lower mobility for Fe:$MoS_2$ compared to $MoS_2$. Despite all the studies on Fe:$MoS_2$, the thermal transport properties of Fe:$MoS_2$ has not been extensively investigated. The findings could be used to design Fe doped $MoS_2$ electronic devices in the future.

In this work, we report the in-plane thermal conductivity of 1L, 2L, and 3L Fe:$MoS_2$ and the interfacial thermal conductance between the Fe:$MoS_2$ and $SiO_2$ substrate for the first time. The opto-thermal Raman technique is employed to measure the thermal properties[27-28] of Fe:$MoS_2$.

This technique is contactless, with no fabrication requirement and capable of directly measuring thermal conductivity without changing the physical properties of pristine samples. It is particularly ideal for measuring 2D materials due to their sub-nanometer thickness. The result has been confirmed by the computational method, density theory function.

**Results**

$MoS_2$ monolayers were synthesized via Low pressure chemical vapor deposition (LPCVD). The schematic of growth using the chemical vapor deposition is shown in Figure 1, and the details are described in the Methods section.

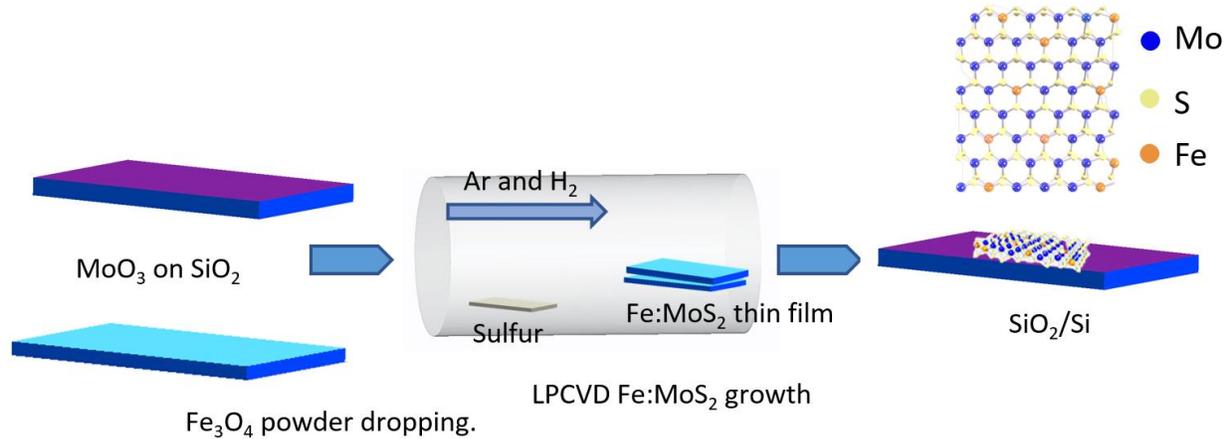

**Figure 1.** Schematic of chemical vapor deposition (CVD) growth process for Fe:$MoS_2$.

The morphology of the Fe:$MoS_2$ samples was characterized by optical microscopy and scanning electron microscopy as shown in Figure 2a and Figure 2b, respectively. To confirm the growth of 1-3L Fe:$MoS_2$ domains, the sample is characterized by the atomic force microscopy (AFM) peak force mode (Bruker Dimension) as shown in Figure 2c. The thicknesses of 1L, 2L, and 3L Fe:$MoS_2$ are 0.8 nm, 1.6 nm, and 2.4 nm. The AFM image also confirms that there is no $Fe_3O_4$ residual remaining on the surface of Fe:$MoS_2$ after wet cleaning and thermal annealing. Figure 2d is the top view schematic of crystal structure of Fe:$MoS_2$.

Raman and photoluminescence (PL) spectroscopy provide evidence that Fe is incorporated into the lattice (Figure 2). To demonstrate the success of doping $MoS_2$ with Fe, the photoluminescence (PL) spectra of the 1L $MoS_2$ and Fe:$MoS_2$ were characterized under a 405 nm laser excitation with 5 mW power at 300 K. There are two broad bands in the PL spectrum associated with two excitons peaks formed by spin-orbit coupling in the valence band (direct excitonic transition). The left PL peak at 1.86 eV represents $MoS_2$ and the right peak at 1.49 eV represents Fe in Fe:$MoS_2$. The red-shift of 0.37 eV caused by defect in $MoS_2$ lattice structure. Figure 2e shows that the PL intensity of the Fe:$MoS_2$ sample dropped significantly compared to the $MoS_2$ sample. This optical quenching results from the $MoS_2$ lattice distortion due to the introducing of the Fe dopants into the lattice and the existence of additional nonradiative channels known as trap states[11, 29].

Due to the highly mobile charges and wide access to shallow trap states, $MoS_2$ is expected to have more electron concentration when Fe dopants are introduced. Fe is an n-type dopant, so Fe-doping is expected to increase electron concentrations in Fe:$MoS_2$ monolayers. Due to the reduction in neutral excitons and the increase in negative trion emission, the peak position of the

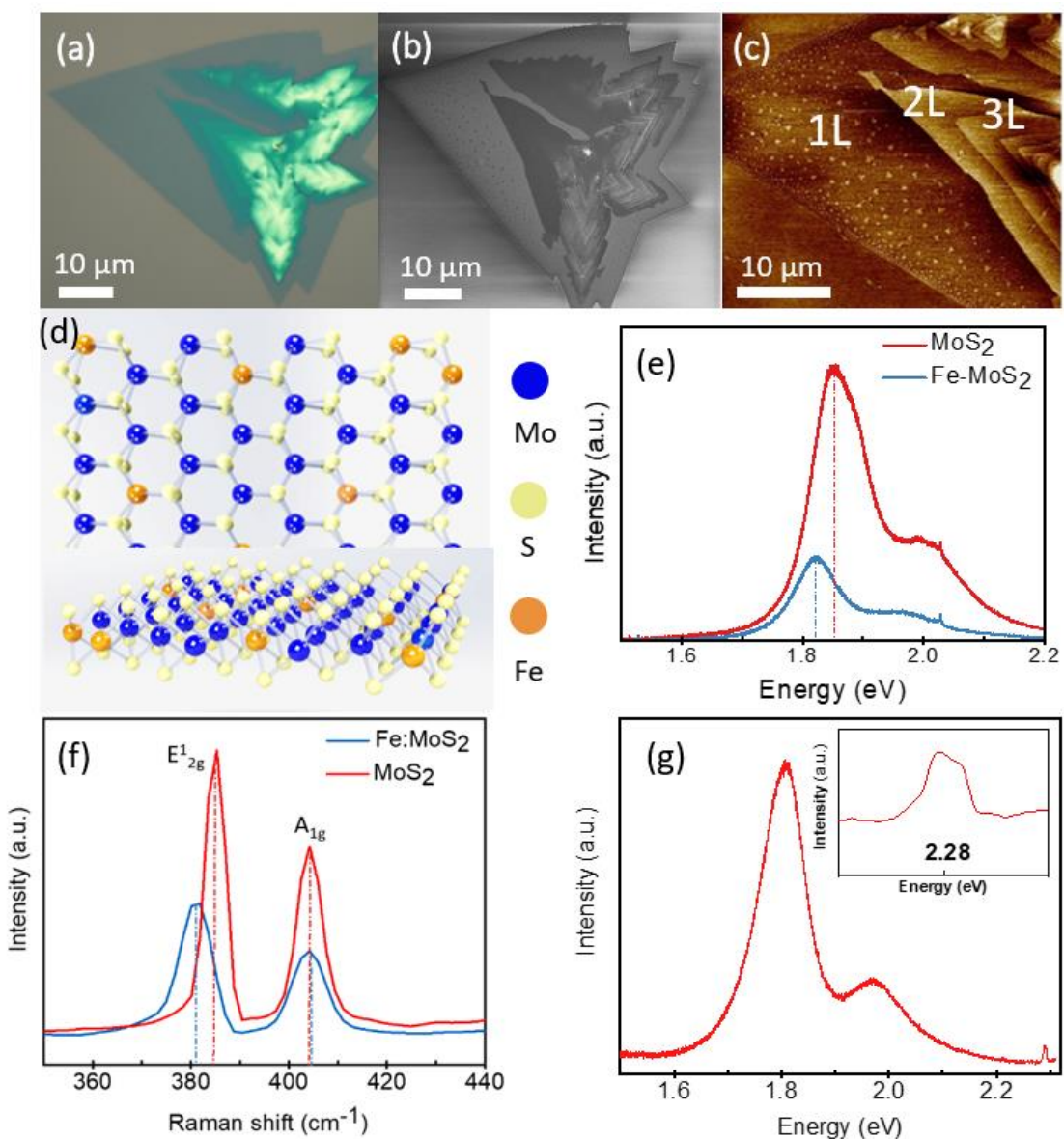

**Figure 2.** (a) Optical microscopy image of 1–3L Fe:MoS$_2$. (b) Scanning electron microscopy image of 1-3L Fe:MoS$_2$. (c) Atomic force microscopy image of 1–3L Fe:MoS$_2$. (d) Top view schematic of crystal structure of Fe:MoS$_2$. (e) Photoluminescent spectra and (f) Raman spectra comparing MoS$_2$ and Fe:MoS$_2$ monolayers. (g) PL spectra of 1L Fe:MoS$_2$ at 300 K. Only Fe:MoS$_2$ monolayers exhibit Fe-related emission (2.28 eV), whereas undoped MoS$_2$ with Fe and Fe$_3$O$_4$ clusters on top does not. This finding suggests that only substitutional incorporation of Fe can produce this transition.

PL would be redshifted, as previous studies demonstrated[30]. The spectra of 1L MoS$_2$ and 1L Fe:MoS$_2$ samples were collected using a micro-Raman spectrometer (Renishaw InVia Raman Microscope system) under 514 nm laser excitation. Both spectra demonstrate the in-plane $E^1_{2g}$ and out-of-plane $A_{1g}$ vibration modes of MoS$_2$ as shown in Figure 2f. The $E^1_{2g}$ and $A_{1g}$ models of the pristine MoS$_2$ sample are at 385.4 cm$^{-1}$ and 404.6 cm$^{-1}$, respectively. The $A_{1g}$ model of the Fe:MoS$_2$ is at 404.2 cm$^{-1}$, which is like that of the pristine MoS$_2$. However, the $E^1_{2g}$ model of the Fe:MoS$_2$ is at 381.1 cm$^{-1}$, indicating a red shift as compared with that of the MoS$_2$. The observed red-shift can be related to the defect in MoS$_2$ lattice structure[16]. In addition, Raman spectroscopy was employed to further confirm that Fe incorporates into the MoS$_2$ lattice. Figure 2g presents a comparison of PL emission for MoS$_2$ and Fe:MoS$_2$. It is clear from this figure that the Fe-related emission at 2.28 eV can only be observed from 1L Fe:MoS$_2$[16], which also validates the substitution of Mo by Fe atoms and there is no peak at 2.28 eV for 1L MoS$_2$. This peak is only observable for 1L Fe:MoS$_2$ at room temperature since 2.28 eV peak is considered as PL emission and is affected by band gap. The band gap changes from direct to indirect as the layer number increases from monolayer to thicker samples. Therefore, the quench for the 2.28 eV peak disappears for samples thicker than 1L[31].

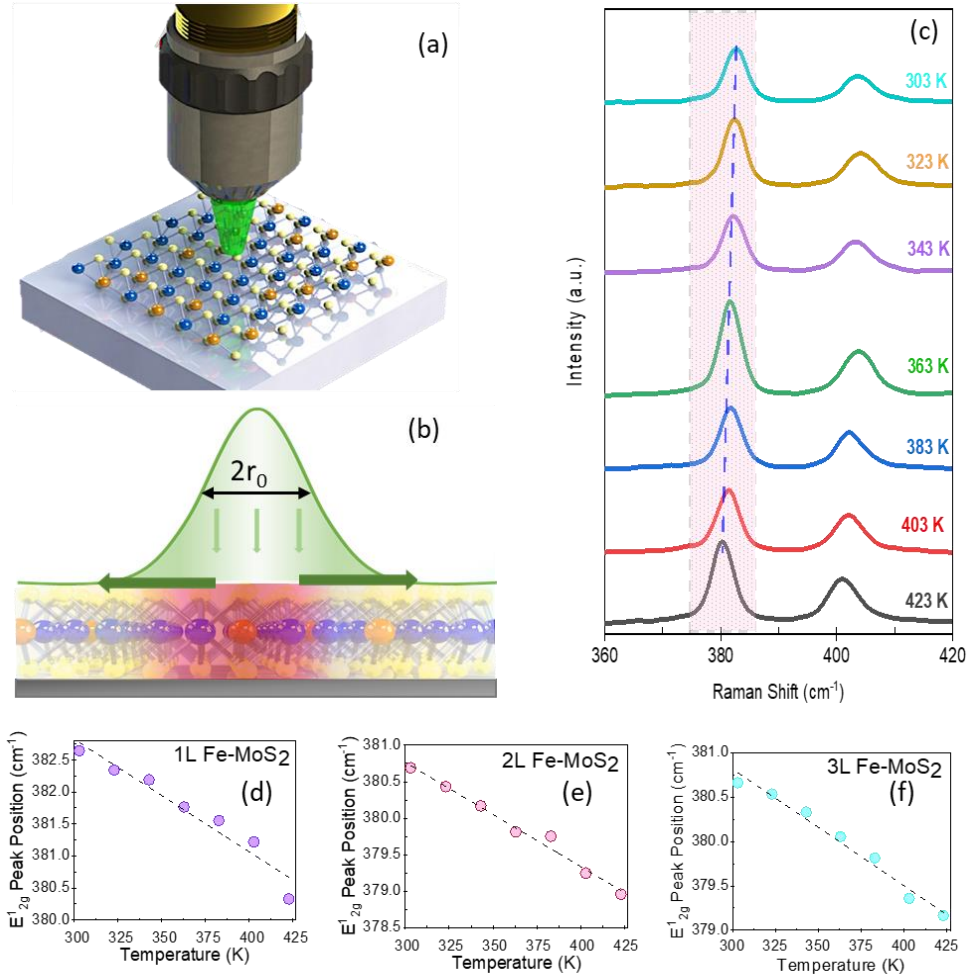

**Figure 3.** **(a)** Schematic of the experimental setup for the Fe:MoS$_2$ sample by the opto-thermal Raman technique. **(b)** Schematic of the Raman laser and thermal transport profiles. **(c)** Raman spectra of 1L Fe:MoS$_2$ recorded at temperatures from 303 K to 423 K. **(d)** The temperature-dependent E$^1_{2g}$ Raman peak shift measured on the 1L Fe:MoS$_2$, **(e)** 2L Fe:MoS$_2$, and **(f)** 3L Fe:MoS$_2$.

Figure 3a presents a schematic of the experimental setup for the Fe:MoS$_2$ sample by the opto-thermal Raman technique, while in Figure 3b a schematic of the Raman laser and thermal transport profiles can be observed. At temperatures ranging from 303 K to 423 K, we calibrated the Raman E$^1_{2g}$ peak position shift rate. A laser with a wavelength λ of 514 nm was used. A Raman spectrum of 1L Fe:MoS$_2$ with temperature dependence is shown in Figure 3c.

To acquire $\chi_P$ (power-dependent coefficient), the E$^1_{2g}$ mode shift of the Fe:MoS$_2$ samples was also measured using the micro-Raman system under the 514 nm laser excitation. Then the power dependent coefficient can be derived from the equation of $\omega(P) = \omega_0 + \chi_P P$, where $\omega(P)$ is the Raman peak frequency at absorbed laser power P, $\omega_0$ is the Raman peak frequency at power 0 and P is the absorbed laser power. The absorbed laser power can be calculated by multiplying the incident laser power with the absorbance ($\alpha$) of the samples. The confocal absorption spectra were recorded with a Nikon Eclipse TE2 equipped with an Acton 500 spectrograph and a PIXIS 100 CCD camera (NT&C Consulting, Germany). Figure 4a shows the absorbance of the 1L, 2L, and 3L Fe:MoS$_2$ as a function of wavelength. We used the interpolated method to extract the absorbance at 514 nm wavelength. The absorbance values of 1L, 2L, and 3L Fe:MoS$_2$ are 0.0127, 0.0344, and 0.0831, respectively. 1L-3L Fe:MoS$_2$ were prepared on a transparent quartz substrate, and the transmittance $T$ of 1L-3L Fe:MoS$_2$ at the wavelength of 480 nm – 800 nm were characterized. The absorbance $A$ were calculated as $A = log10\,(1/T)$. The E$^1_{2g}$ peak position as a function of the absorbed laser power was characterized using both 100× and 50× objective lens of the Raman system as shown in Figure 4b, c, and d. The laser light acted as both the thermometer and the heat source for the Fe:MoS$_2$ flakes. It can be observed that the E$^1_{2g}$ peak of all the samples decreases almost linearly as a function of the absorbed laser power. This is because the local temperature rise induced by the laser heating resulted in red shift due to the phonon softening effect. The convection through the air is negligible since it is less than 0.01% for all the measured samples[18]. Both $\chi_T$ and $\chi_P$ are extracted via linear regression and summarized in Table 1.

| | Temperature coefficient (cm$^{-1}$/K) | Absorbance (%) | Absorbed Power Shift Rate (cm$^{-1}$/μw) | |
|---|---|---|---|---|
| | | | 0.18 μm spot size | 0.31 μm spot size |
| 1L Fe:MoS$_2$ | $-0.0176 \pm 0.0020$ | 1.27 | $-0.1185 \pm 0.0036$ | $-0.0644 \pm 0.00150$ |
| 2L Fe:MoS$_2$ | $-0.0143 \pm 0.0008$ | 3.44 | $-0.0383 \pm 0.0009$ | $-0.0182 \pm 0.00004$ |
| 3L Fe:MoS$_2$ | $-0.0131 \pm 0.0008$ | 8.31 | $-0.0158 \pm 0.0004$ | $-0.0067 \pm 0.00004$ |

**Table 1.** First-Order Temperature Coefficients, Absorbance, and Power Shift Rates of 1-3L Fe:MoS$_2$.

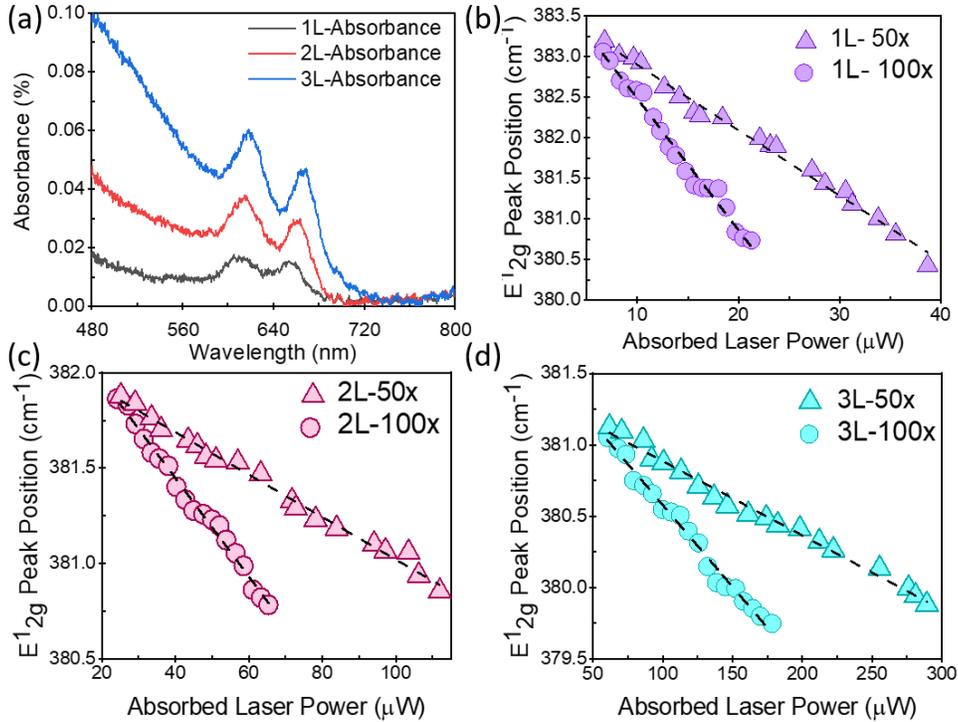

**Figure 4.** (a) Absorbance of 1-3L Fe:MoS$_2$ at wavelengths from 480 nm to 800 nm. (b) Power-dependent E$_{2g}^1$ Raman peak shift characterized using two different laser spot sizes, on the 1L Fe:MoS$_2$, (c) 2L Fe:MoS$_2$, and (d) 3L Fe:MoS$_2$.

To calculate the thermal conductivity, one also need to know the laser spot size of the Raman system. The laser spot size of both 100× and 50× objective lenses was measured using the knife-edge method[32]. In this method, the laser moves across a sharp edge and using the Raman mapping feature the Si peak is monitored. The Si peak intensity as a function of position had been studied with Gaussian distribution as mentioned in equation below

$$I = I_0 \exp\exp\left(\frac{(x-x_c)^2}{r_0^2}\right) \tag{1}$$

where $r_0$ is the laser spot size, $x_c$ is the center of peak position, and $I_0$ is the initial intensity. The laser spot size passing through 100x and 50x was 0.18 µm and 0.31 µm, respectively. The measurements were made in the air at ambient temperature. Laser spot size can also be estimated from the numerical aperture. $r_0 = \frac{\lambda}{\pi.NA}$ where $\lambda$ is the laser wavelength and NA is the numerical aperture. For 100x and 50x, the calculated result using the numerical aperture is 0.19 µm and 0.22 µm, respectively. The substrate interaction can make the mean free paths of phonons in Fe:MoS$_2$ smaller than the laser beam size, especially for long-wavelength phonons[33]. In the cylindrical coordinate T(r), the temperature distribution is related to the in-plane thermal conductivity, interfacial thermal conductance between the flake and substrate, and the absorbed laser power. The volumetric laser heating power density q'''(r) based on the Gaussian profile of the laser equation below is

$$q'''(r) = \frac{P}{t.\pi r_0^2} \exp\exp\left(\frac{-r^2}{r_0^2}\right) \tag{2}$$

where P is the absorbed laser power, t is the thickness of the flake, and $r_0$ is the laser spot size. After taking the interfacial thermal conductance between the flake and the substrate into consideration, T(r) is presented by the equation below

$$\frac{1}{r}\frac{d}{dr}\left(r\frac{dT(r)}{dr}\right) - \frac{g}{k_s t}(T(r) - T_a) + \frac{q'''(r)}{k_s} = 0 \tag{3}$$

where g is the interfacial thermal conductance between the flake and the substrate, $K_s$ is the in-plane thermal conductivity of Fe:MoS$_2$ and $T_a$ is the ambient temperature. The weighted result can be calculated by considering the Gaussian profile of the laser spot. The equation below is the average temperature.

$$T_m = \frac{\int_0^\infty T(r)\exp\left(\frac{-r^2}{r_0^2}\right)rdr}{\int_0^\infty \exp\left(\frac{-r^2}{r_0^2}\right)rdr} \tag{4}$$

To solve the equations, boundary conditions need to be defined. Based on the Gaussian profile of the laser $\frac{dT}{dr}|_{r=0} = 0$, the maximum temperature is in the center where the laser shines on the sample ($T(r = 0) = T_{max}$ and further away from the center, the temperature is the ambient temperature $T(r \to \infty) = T_a$. We also need to calculate the thermal resistance of Fe:MoS$_2$, which can be calculated as $R_m = \frac{T_m}{P}$ and relate the thermal resistance to the thermal conductivity and the interfacial thermal conductance between the flake and the substrate. The ratio of two $R_m$ characterized from two laser spot sizes is a function of $g/K_s$. The calculated results are $K_s = 24 \pm 11$ W/m·K, and g= 0.3 ± 0.2 MW/m$^2$·K for 1L Fe:MoS$_2$, $K_s = 18 \pm 9$ W/m·K, and g= 1.1 ± 0.7 MW/m$^2$·K for 2L Fe:MoS$_2$, and $K_s = 16 \pm 8$ W/m·K, and g = 3.0 ± 2.3 MW/m$^2$·K for 3L Fe:MoS$_2$.

To obtain the relationship between Raman peak shift and temperature rise, we also heated the samples from room temperature to 500 K on the heating platform. The optothermal Raman characterizations were conducted with a Raman laser of low power (100 µW). The thermal conductivity and interfacial thermal conductance between the flake and substrate are summarized in Table 2.

|  | Thermal Conductivity (W/m·K) | | Interfacial Thermal Conductance (MW/m²·K) |
| --- | --- | --- | --- |
|  | At 300 K | At 500 K |  |
| 1L Fe:MoS$_2$ | 24 ± 11 | 19 ± 10 | 0.3 ± 0.2 |
| 2L Fe:MoS$_2$ | 18 ± 9 | 11 ± 6 | 1.1 ± 0.7 |
| 3L Fe:MoS$_2$ | 16 ± 8 | 8 ± 3 | 3.0 ± 2.3 |

**Table 2.** Thermal Conductivities of 1–3L Fe:MoS$_2$ at room temperature (300K) and at 500K, and their Interfacial Thermal Conductance.

**Discussion**

Thermal conductivities of 1-3L Fe:MoS$_2$ have been explored and found in the range of 16 – 24 W/m·K which is about 40% lower than the reported values for MoS$_2$[18]. This value is also smaller than other TMDC materials, such as MoS$_2$, MoSe$_2$, and WSe$_2$, due to the imperfect lattice structure of MoS$_2$ from the doping engineering. The iron doped MoS$_2$ causes imperfection in the lattice structure, leading to phonon scattering and a further reduction in thermal conductivity. The phonon scattering rate caused by imperfections can be derived from Green's function approach or calculated using Fermi's golden rule[34-35]. The phonon scattering rate for 2D materials can be estimated by the formula $\frac{1}{\tau_{qs}^D} \propto \left(\frac{\omega^3}{v^2}\right) \sum_i f_i \left[(1 - \frac{m_i}{m})^2 + \varepsilon(\gamma (1 - \frac{r_{a,i}}{r_a})^2\right]$, where m and r$_a$ are the average atom mass and radius, $v$ is the phonon group velocity, $\gamma$ is the Gruneisen parameter, m$_i$, r$_a$, i, and f$_i$ are the mass, radius, and fraction concentration of type i atom, and $\varepsilon$ is a parameter for defect scattering due to lattice distortion and mass difference. This expression indicates the likelihood of high-frequency phonon modes scattering by defects. Since the Fe atoms substitute the Mo atoms in MoS$_2$ lattice structure, the difference between the atomic radius and the atomic mass of these two materials cause a change in phonon scattering rate[36].

A decreasing trend of thermal conductivities and an increasing trend in interfacial thermal conductance have been observed with increasing layer numbers of Fe:MoS$_2$. The decreasing trend of thermal conductivities from thinner layer to thicker layer is due to the stronger phonon-boundary scattering. The strength of boundary condition scattering, which is defined as $\Gamma_{qs}^B \propto \frac{|v_{qs}^z|}{t}$ where t is the thickness, decreases in the layered 2D crystals due to the weak van der Waals interaction [36]. In addition, the decreasing trend is associated with the Umkkalpp phonon scattering, crystal's phonon anharmonicity, and reduction of ballistic resistance that affects phonon mean free path for thicker flakes. Other studies on the MoS$_2$, MoSe$_2$, WSe$_2$, and hBN also faced the decreasing trend of thermal conductivity with increasing the number of layers[18,37-38]. The increasing trend in interfacial thermal conductance could be explained that as the number of layers increases, the interface between 2D materials and the substrate becomes more tightly coupled. In addition, there are less defects in 2D materials with increased layers because more layers can act as "bridge" between 2D materials and the substrate. We also compared to the undoped CVD MoS$_2$ that we synthesized with the Fe doped CVD MoS$_2$ in this work(with Fe concentration 0.3-0.5%), and found the Fe doped CVD MoS$_2$'s thermal conductivity is ~30% smaller. This decrease in thermal conductivity is due to the defects in the material created during the doping process.

Computational studies were carried out in parallel to verify the experimental results. we utilized a combination of phonopy, thirdorder, and ShengBTE software, integrated with VASP first-principles DFT calculations, to compute the thermal conductivity of Fe-doped MoS$_2$. The second- and third-order interatomic force constants were extracted and employed to solve the phonon Boltzmann transport equation. For this, we constructed a 4×4×1 supercell of monolayer MoS$_2$, with one Mo atom substituted by an Fe atom to introduce doping. Following structural optimization, the atomic configuration corresponding to the minimum energy was identified. The cutoff energy for plane-wave basis sets was set to 400 eV, ensuring sufficient accuracy, while the energy convergence criterion was set to $1\times10^{-6}$ eV, which provided precise structural optimization. Additionally, we used a Γ-centered k-point mesh of 6×6×1 for Brillouin zone sampling and employed a Methfessel-Paxton smearing method with a width of 0.05 eV to ensure electronic structure convergence during the relaxation process. the cutoff distance was set to 0.3 Å in the calculation of third-order force constants for neighboring atoms. We employed a 10×10×10 q-grid to more accurately estimate phonon frequencies and phonon lifetimes, ensuring better resolution of the phonon characteristics for our calculations.

The thermal conductivity results revealed that for the monolayer Fe-doped MoS$_2$, the in-plane thermal conductivity was 28.3 W/mK. In contrast, for the bilayer Fe-doped MoS$_2$, the in-plane thermal conductivity dropped to 20.3 W/mK. The in-plane thermal conductivity of the 3L material is 18.8 W/mK. The computational results are in good agreement with the experimental findings. These findings indicate that the significant reduction in thermal conductivity for multilayered 2D materials, as compared to their monolayer counterparts, can be attributed to the poor efficiency of interlayer thermal transport. The out-of-plane thermal conductivity remains several orders of magnitude lower than the in-plane thermal conductivity, which is primarily responsible for the observed suppression of thermal transport in the multilayer structure.

**Methods**

MoS$_2$ monolayers were synthesized via Low pressure chemical vapor deposition (LPCVD). Before growth, a thin MoO$_3$ layer was prepared using physical vapor deposition (PVD) of e-beam evaporating MoO$_3$ pellet onto a Si substrate with 300 nm-thick thermal oxides. Subsequently, the MoO$_3$-deposited substrate was in direct contact with a different SiO$_2$/Si substrate. On the SiO$_2$/Si substrate, monolayers of Fe:MoS$_2$ were grown. The following steps were taken to dope the Fe atoms: the MoO$_3$-deposited substrate was rinsed with deionized (DI) water to deposit a thin layer of water on the SiO$_2$ surface to ensure that the Fe$_3$O$_4$ particles were distributed uniformly. Fe$_3$O$_4$ particles were uniformly cast onto the SiO$_2$/Si substrate and then the substrate was baked at 110 °C for 5 minutes. During the growth, the furnace was heated with a ramping rate of 18 °C/min and held for 15 min at 850 °C. Upon the heating procedure, an argon gas (30 sccm) was supplied at 300 °C, and a hydrogen gas (15 sccm) was delivered at 760 °C. Sulfur was supplied when the furnace temperature reached 790 °C. After the growth, a few millimeters of Fe:MoS$_2$ monolayers were obtained.

**Conclusion**

It is the first work to explore the thermal transport properties of the novel magnetic 2D material Fe:MoS$_2$ in its 1L to 3L forms, both experimentally and computationally. Fe:MoS$_2$ is produced via the LPCVD method, with Fe atoms in the lattice structure of Fe:$MoS_2$ characterized by Raman spectroscopy and photoluminescence. Our refined opto-thermal Raman method was used to characterize the in-plane thermal conductivity and interfacial thermal conductance of Fe:$MoS_2$. A decreasing trend has been observed for thermal conductivities of Fe:$MoS_2$ as the layer number increases. The in-plane thermal conductivities of 1L, 2L, and 3L Fe:$MoS_2$ are $24 \pm 11$ W/m·K, $18 \pm 9$ W/m·K, and $16 \pm 8$ W/m·K and the interfacial thermal conductivity are $0.3 \pm 0.2$ MW/$m^2$·K, $1.1 \pm 0.7$ MW/$m^2$·K, and $3.0 \pm 2.3$ MW/$m^2$·K respectively. These values will help us to understand their thermal behavior, the critical factor that has a significant limitation on nano- and quantum devices. Experimental results show a layer-dependent trend of thermal conductivity.

## Data availability

The datasets used and analysed during the current study available from the corresponding author on reasonable request.

## Acknowledgements

This work was supported by the National Science Foundation CAREER Award (Grant CBET-2145417) and LEAPS Award (Grant DMR-2137883).

## Author Contributions

X.Z. contributed to the conceptualization and the methodology. E.E. and X.Z. contributed to the investigation. M.F. and E.Y. contributed to the material synthesis. M.L. contributed to the optical characterization. E.Y. contributed to writing the original draft. X.Z. contributed to writing the review and editing. All authors have given approval to the final version of the manuscript.

## Declarations

### Competing interests

The authors declare no conflict of interest.

## Additional information

### Corresponding Author

*Xian Zhang. Email: xzhang4@stevens.edu

### Funding Sources

National Science Foundation CAREER Award (Grant CBET-2145417) and LEAPS Award (Grant DMR-2137883).